\documentstyle[amssymb,prb,aps,twocolumn,epsf]{revtex}

\begin{document}
\twocolumn[\hsize\textwidth\columnwidth\hsize\csname@twocolumnfalse\endcsname

\title{Transport properties and structures of vortex matter in layered
superconductors}
\author{M. F. Laguna, D. Dom\'{\i}nguez and C. A. Balseiro}
\address{ Centro At\'{o}mico Bariloche and Instituto Balseiro, Comisi\'{o}n Nacional de Energ\'{\i}a
At\'{o}mica, 8400 San Carlos de Bariloche, R\'{\i}o Negro. Argentina}

\maketitle

\begin{abstract}
In this paper we analyze the structure, phase transitions and some transport
properties of the vortex system when the external magnetic field lies
parallel to the planes in layered superconductors. We show that experimental results for resistivity
are qualitatively consistent with numerical simulations that describe the
melting of a commensurate rotated lattice. However for some magnetic fields, the
structure factor indicates the occurrence of smectic peaks at an intermediate
temperature regime.
\end{abstract}


]

\section{INTRODUCTION}

The discovery of high $T_{c\text{ }}$superconductivity renewed the interest
for the thermodynamic, structural and dynamical properties of vortex matter.
Due to the large temperatures available for the vortex system and an
important number of relevant parameters, as anisotropy, disorder and
the magnitude and direction of the external magnetic field, the physics of
these systems is very rich. First and second order phase transitions between
a low temperature solid phase an a high temperature liquid phase have been
predicted and observed experimentally. The low temperature phase can be
either a crystalline or a glassy phase depending on the amount and nature of
the disorder present in the sample. \cite{crabtree,blatter}

All the cuprate high $T_{c\text{ }}$materials have in common a crystalline
structure based on the existence of CuO planes. This makes all of them
anisotropic materials with a layered structure. In the usual convention, the 
$c$-axis points in the direction perpendicular to the planes and the $a$ and 
$b$ axis are the in-plane crystalline axis.

Concerning the study of the vortex properties in these materials, most of
the work was devoted to the study of the configuration in which the external
field points along the $c$-axis, particularly in the case of experimental
work. There is however an increasing interest in the properties of
the system for fields parallel to the planes. \cite
{bula,ivlev,Kwok,bal1,bal2,Hu,Olson} Theoretically, in many cases, the system is simulated
as a highly anisotropic but homogeneous system. \cite{CDK} The planes
however act as a strong potential for the vortices that tend to localize
them between planes where superconductivity is weak.

In a clean system and at low temperatures, for the external field parallel to the $ab-$planes, the
vortices form an anisotropy - distorted Abrikosov lattice. If the lattice is commensurate with the
periodic potential due to the planes, the ground state of the system is a commensurate structure like the
one schematically shown in Fig.(1a). In this particular case, the vortex
density $\delta$ in the direction perpendicular to the planes is modulated with a
period given by the lattice parameter of the vortex lattice. For this type
of structures, the most salient theoretical prediction is the existence, at
an intermediate temperature interval, of a smectic phase.\cite{bal1,bal2} In
this picture the transition between the crystalline state and the high
temperature liquid state takes place in two steps. In the intermediate
regime the system develops long (or quasi-long) range order in the
direction perpendicular to the planes, \textit{i.e.} the liquid develops
some density oscillations along this direction as a precursor of the frozen
state. This phase is known as the smectic phase. The existence of the
smectic phase is unconfirmed and there are many relevant questions
concerning its nature and stability. One of them is what happens if the
field does not produce a commensurate structure like the one shown in the
figure. One possibility is that the lattice locally retains its structure
and orientation and generates discommensuration to accommodate to the
periodic potential generated by the planes. If the mismatch is not too large
the discommensurations are far apart and the physical properties of the
system are not very affected. In this case it could be possible to detect
the existence of the smectic phase without tuning the external field to its
commensurate value. The other alternative is that as the field is changed,
the vortex lattice rotates and distorts - due to the anisotropic properties
of the system- to form a new structure that is commensurate with the
periodic potential like the one shown in Fig.(1b). In this paper we will
show that at low fields, as the field changes, between two consecutive
commensurate structures of the type of Fig.(1a), there are a number of
commensurate phases of the type illustrated in Fig.(1b).

For the particular case of the rotated structure shown in the figure, the
vortex density $\delta$ of the lattice is the same between any two consecutive planes. In
this case, as the temperature increases, the lattice could melt going directly from the
solid to the liquid state. If this is the case, it would be very difficult
to observe the smectic phase for an arbitrary chosen external field.

The transport properties of the vortex system for the configuration of
interest were recently measured\cite{Grig} and the data were compared with
the theory of the smectic phase. Experiments and theory are only partially
consistent.

In the rest of the paper we discuss, in terms of the London theory with a
periodic potential, the stability of the different commensurate phases. We
also present some numerical simulations for the resistivity in layered
structures and analyze the corresponding low temperature vortex configurations.
We will show that the experimental results for the resistivity are
qualitatively consistent with numerical simulations. For some fields, the
structure factor indicates the occurrence of smectic peaks at an
intermediate temperature regime.

\begin{figure}[tbp]
\centerline{\epsfxsize=8.5cm \epsfbox{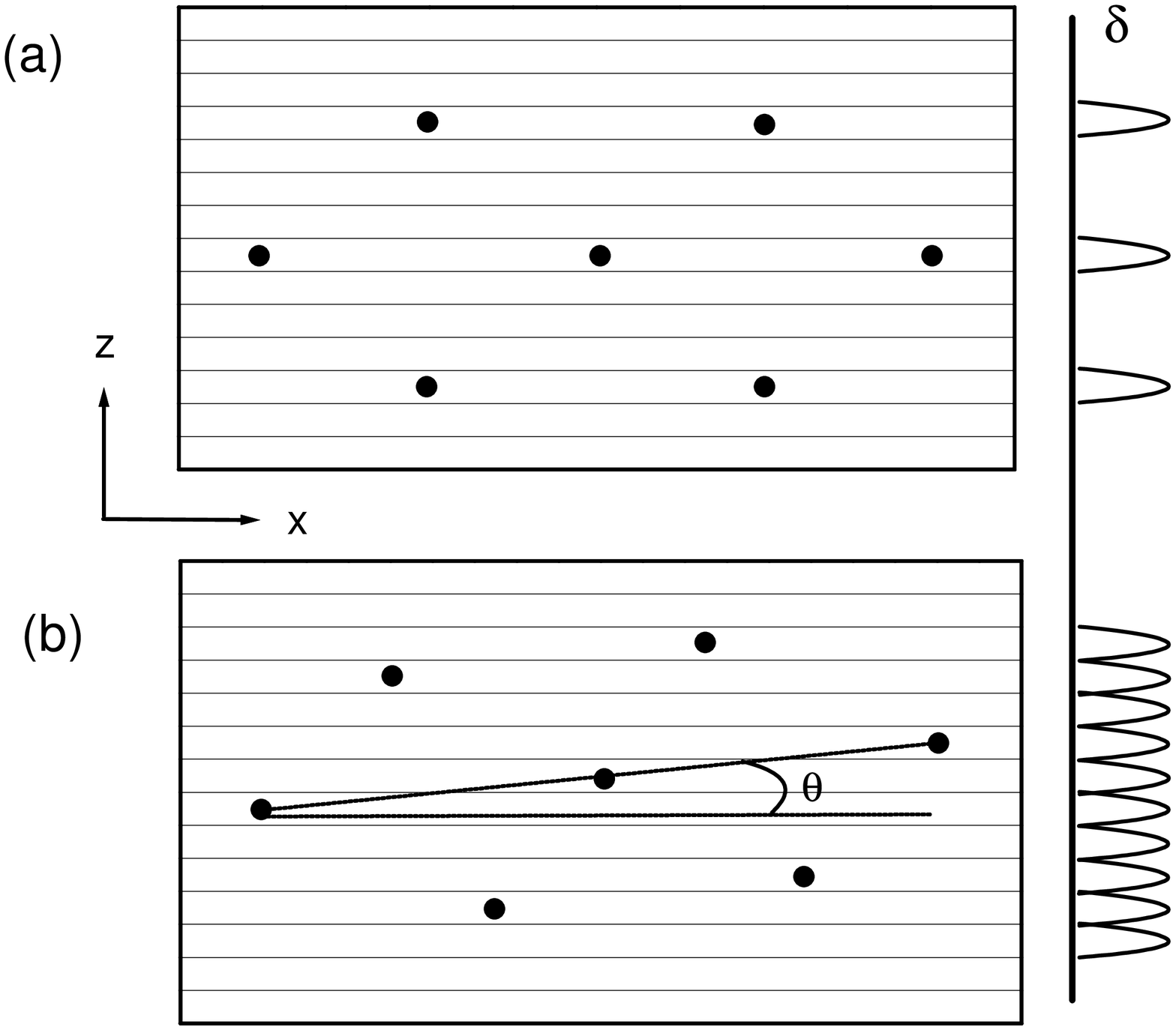}}
\caption{Commensurate structures at low temperatures. In (a) the magnetic field
applied in the $y$-direction generates a non-rotated vortex lattice for which the
vortex density $\delta$ is modulated in the direction perpendicular to the
ab-planes. In (b) the vortex lattice is rotated an angle $\theta$, and $\delta$ is the same for all
the planes.}
\label{fig1}
\end{figure}

\section{COMMENSURATE STATES IN THE LONDON APPROXIMATION}

In order to analyze the low temperature structures of the vortex lattice we
resort to the London approach in an anisotropic material and in the presence
of a uniaxial potential representing the effect of the planes. Following the
work by Campbell, Doria and Kogan\cite{CDK} (CDK), the free energy per unit
length in the direction of vortices for an anisotropic material is given by: 
\begin{equation}
F_{0}=\int ({\mathbf h}^{2}+\lambda ^{2}m_{ij}\hspace{0.04in}curl_{i}{\mathbf h%
\hspace{0.04in}}curl_{j}{\mathbf h)}\frac{dxdz}{8\pi }  \label{f0}
\end{equation}
where ${\mathbf h}(x,z)$ is the local magnetic field in the plane
perpendicular to the vortices, $\lambda ^{2}$ is proportional to the \textit{%
average mass} $M_{av}=(M_{1}M_{2}M_{3})^{1/3}$ with $M_{k}$
being the principal values of the mass tensor $M_{ij}$ and $%
m_{ij}=M_{ij}/M_{av}\hspace{0.04in}$is the \textit{effective-mass tensor}.
We consider the external field in the $y-$direction (parallel to the planes)
that coincides with one of the principal axis of the crystal, ${\mathbf h}=(0,h_{y},0)$. Then we take $%
m_{xx}=m_{yy}=m_{1}$ and $m_{zz}=m_{3}$.

For a vortex lattice, $h_{y}(x,z)$ is a periodic function
with nonzero Fourier components $h_{y}\mathbf{(G})$, where $\mathbf{%
G}$ are the reciprocal lattice vectors of the vortex lattice. The free
energy is minimized with respect to $\mathbf{h}$ and following CDK, $F_{0}$
is given by:

\begin{equation}
F_{0}=\frac{B^{2}}{8\pi }\sum_{\mathbf{G}}\frac{1}{1+\lambda
^{2}(m_{1}G_{x}+m_{3}G_{z})}  ,\label{fG}
\end{equation}
where $B=\phi _{0}n$ is the magnetic induction, $\phi _{0}$ is the
flux quantum and $n$ is the number density of vortices. In this expression,
the summation is over all the reciprocal lattice vectors $\mathbf{G}$ of the
vortex lattice. The mass anisotropy distorts the hexagonal lattice
compressing it in the $z-$direction and expanding it along the $x-$%
direction. The lattice can rotate to form an angle $\theta $ with the $x-$%
axis as shown in Fig (1b). For this particular configuration, in which the
magnetic field is parallel to one of the principal axis of the crystal, CDK shown that the $%
F_{0}$ contribution to the free energy does not depend on $\theta $.

The above considerations are valid for continuous anisotropic superconductors. In layered superconductors, the
presence of planes introduces a periodic potential that will partially break the degeneracy in $\theta $.
In a first order approximation, we describe the effect of the planes by including a periodic potential that
tends to localize the vortices in the interplane spacing:

\begin{equation}
F=F_{0}+\int h_{y}(x,z)V(z)dxdz  \label{f}
\end{equation}
where $V(z)$ is a periodic potential with the periodicity of the $c-$axis
lattice parameter.
The condition for commensurability of the vortex lattice with the periodic potential, corresponding to
the situation in which all vortices are placed at a minimum of $V(z)$, is
given by the condition $x_{j}=ns$ where $x_{j}$ is the $x$-component of the
coordinate of vortex $j$, $n$ is an integer and $s$ is the $c$-axis lattice
parameter (the interplane distance). In the reciprocal space this condition gives a relation between
the reciprocal lattice vectors and the vector ${\mathbf Q}$ of the periodic potential $V(z)$:
\begin{equation}
{\mathbf Q}=\frac{2\pi}{s} {\bf \hat{z}} = p {\bf G_{1}} + q {\bf G_{2}} 
\end{equation}
where $p$ and $q$ are integers and  
\begin{equation}
{\bf G_{1}}= \frac{2 \pi}{\gamma L} \frac{\sin (\theta +\pi /3)}{\sin \pi /3} {\bf \hat{x}} - 
\frac{2 \pi \gamma}{L} \frac{\cos(\theta +\pi /3)}{\cos \pi /3} {\bf \hat{y}}
\end{equation}
\begin{equation}
{\bf G_{2}}= -\frac{2 \pi}{\gamma L} \frac{\sin \theta}{\sin \pi /3} {\bf \hat{x}} +
\frac{2 \pi \gamma}{L} \frac{\cos \theta}{\cos \pi /3} {\bf \hat{y}}
\end{equation}
with $\gamma ^{2}=\sqrt{m_{1}/m_{3}}$ the anisotropy factor and $L^{2}= (2 \Phi_{0})/(\sqrt{3}B)$. 

This condition can be put in the form: 
\begin{equation}
\tan (\theta )=\frac{p\cos (\pi /3)-q}{p\sin (\pi /3)}  \label{alfa}
\end{equation}
and

\begin{equation}
B=\frac{2\phi _{0}\gamma ^{2}\sin ^{2}(\pi /3)}{\sqrt{3}s^{2}(p\sin (\theta
+\pi /3)-q\sin (\theta ))^{2}}  \label{b}
\end{equation}
In Fig.(2), all angles $0\leqslant \theta \leqslant \pi /6
$ satisfying the condition for commensurability are shown as a function of
the magnetic induction $B$. In the London approximation, since the free
energy is $\theta $ independent, all commensurate states obtained for a
given magnetic induction $B$ are degenerate. 
As can be seen in Fig.(2), for low fields, between two consecutive commensurate structures of the type of
Fig.(1a) - corresponding to $\theta=0$ or $\pi/6$ - there are a large number of commensurate phases with $\theta
\neq 0$. As the field increases the number of commensurate phases decreases and in the limit of high magnetic
fields (or
high anisotropy) there are only two undistorted lattices corresponding to
$\theta=0$ and $\pi/6$, and at all temperatures the vortex density is the same
between any two consecutive planes. In this limit, increasing the magnetic field
compresses the vortex lattice in the $z-$direction. Our simple model is not
appropriate to describe this limit.\cite{hu2,ivlev2}

\begin{figure}[tbp]
\centerline{\epsfxsize=8.5cm \epsfbox{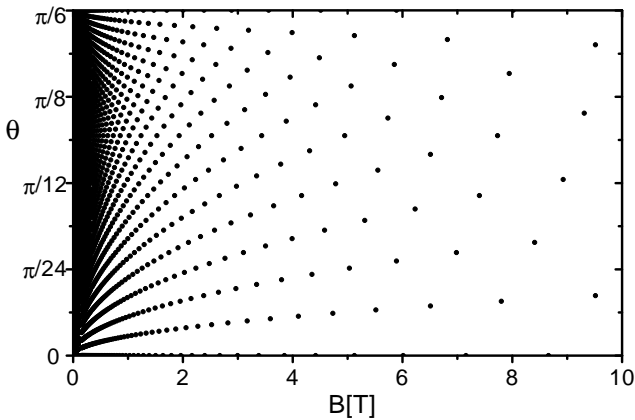}}
\caption{Angles of rotation of the vortex lattice vs. magnetic field. Each point
of the diagram correponds to a commensurate state of minimum energy. Magnetic
field is in Tesla and the parameters are those typically used
to describe YBaCuO: $ s = 10\AA $ and mass anisotropy $m_{1}/m_{3}=50$.}
\label{fig2}
\end{figure}

For all angles, the high
anisotropy generates structures where the vortex-vortex distance along the
plane directions is much larger than the distance perpendicular to the
planes, giving rise to vortex chains.
Recently Hu and Tachiki\cite{Hu} studied the ground
state configuration of the vortex lattice, using Monte Carlo
simulations in the highly anisotropic XY model for large systems. They found structures
consisting 
of buckling vortex chains. We interpret these results as rotated lattices
with discommensurations due to the mismatch of the vortex lattice parameter
with the interplane distance. In the numerical simulations, transverse
discretization of the space could be another source of discommensurations.
Although in general, we expect the discommensuration to form an angle of $%
45^{0}$ with respect to the planes,\cite{prokov,martin} the boundary
conditions in a finite sample could stabilize discommensurations parallel to
the planes. The buckling vortex chains structures are consistent with
Bitter-pattern observations, however in the experimental case twin
boundaries may be relevant to stabilize the observed structure.\cite
{dolan,ivlev}

\section{NUMERICAL SIMULATIONS}

In this section we present numerical simulations for the transport
properties and structure factor in an anisotropic system described by a
three-dimensional (3D) Josephson-junction array. The model has been
extensively used and is described in Refs.[17,18] and here we give
only a brief summary of the method.

The equilibrium physics of this system is described by the Hamiltonian of a three-dimensional
frustrated XY model:\cite{hetzel,li,hu3}
\begin{eqnarray}
H &=& - E_{J}\sum_{<i,i^{\prime }> \in ab-plane} \cos(\varphi ^{i}-\varphi ^{i^{\prime}}-A^{ii^{\prime
 }}) +\nonumber\\
& &\frac{1}{\gamma ^{2}} \sum_{<i,i^{\prime }> \in c-axis}\cos(\varphi ^{i}-\varphi ^{i^{\prime}}-A^{ii^{\prime
 }})
\end{eqnarray}
with $E_{J}= \frac{\phi _{0}^{2} s}{16 \pi^{3} \lambda ^{2}}$ ($\lambda$ is the penetration
length in the $ab$ planes and $s$ is the interplane distance) and $A^{ii^{\prime }}\equiv (2e/\hbar c)
\int_{i}^{i^{\prime}} {\bf A}.d{\bf l}$ the integral of the vector potential of the external magnetic
field from site $i$ to site $i^{\prime}$. The phases $\varphi ^{i}(t)$ are defined in the nodes of
a lattice and represent the phase of the order parameter. The thermodynamics of this Hamiltonian
coincides with the equilibrium properties of the 3D Josephson-junction array.

The dynamics of the 3D Josephson-junction array is contained in the time
evolution of the phases $\varphi ^{i}(t)$. Nearest neighbor
nodes are coupled with Josephson junctions characterized by critical
currents $I_{c}^{ii^{\prime }}$ and normal resistances $R^{ii^{\prime }}$.
The equations describing the model are

\begin{eqnarray}
j^{ii^{\prime }}&=&I_{c}^{ii^{\prime }}\sin (\varphi ^{i}-\varphi ^{i^{\prime
}}-A^{ii^{\prime }})+\nonumber\\
& &\frac{\hbar }{2eR^{ii^{\prime }}}\frac{\partial
(\varphi ^{i}-\varphi ^{i^{\prime }})}{\partial t}+\eta ^{ii^{\prime }}(t)
\label{jii}
\end{eqnarray}

\begin{equation}
\sum_{\{i^{\prime }\}}j^{ii^{\prime }}=j_{ext}^{i}  \label{sj}
\end{equation}

Equation (\ref{jii}) gives the current between the nearest neighbor nodes $i$
and $i^{\prime }$ with phases $\varphi ^{i}$ and $\varphi ^{i^{\prime }}$. Here $\eta ^{ii^{\prime }}(t)$ is an
uncorrelated gaussian noise that incorporates the effect of the temperature. Equation (\ref{sj}) ensures the
current conservation at each node and $j_{ext}^{i}$ is the external current
applied at node $i$. Equations (\ref{jii}) and (\ref{sj}) are numerically
integrated on time using a Runge-Kutta method. The typical time step used is 
$\Delta t=.1\tau _{J}$ ($\tau _{J}=\phi _{0}/2\pi R_{0}I_{c}$) and the number of iterations is $20,000
\leq N \leq 100,000$. Details of the numerical
method have been presented in previous works.\cite{Mingo,jagla}
The mean in-plane
critical currents $I_{c}^{\parallel}$ are larger than the mean inter-plane critical
currents $I_{c}^{\perp}$ by an anisotropy factor $\gamma ^{2}$ ( $\gamma
^{2}\equiv I_{c}^{\parallel}/I_{c}^{\perp }$ ), and $I_{c}^{\parallel}=I_{c}=\frac{2 \pi E_{J}}{\phi_{0}}$.
At the same time, the ratio between
the in-plane resistance $R^{\parallel}$ and the out of plane resistance $R^{\perp }$
is given by $1/\gamma ^{2}$. 
We also consider a small amount of disorder by taking a
uniform distribution of critical currents of width $\Delta$ defined as

\begin{equation}
\Delta = \frac{(I_{c} ^{max} - I_{c} ^{min})} {(I_{c} ^{max} + I_{c} ^{min})}
\end{equation}

where $I_{c} ^{max}$ and $I_{c} ^{min}$ are the maximum and minimum values of the critical current in the
corresponding directions. The disorder simulated are typically $0 \leq \Delta \leq .1$. 
As in the previous section, we take the $z-$%
direction as the direction perpendicular to the planes (parallel to the $c$%
-axis of the crystal) and the external magnetic field along the $y-$%
direction. The magnetic fields simulated are $ 1/72 \leq f \leq 1/6$, where $f=B a^{2}/\phi_{0}$ ($a$ is
the square lattice period). 
The values of anisotropy are $1 \leq \gamma ^{2} \leq 45$ and typical system sizes are
$8 \leq L_{x},L_{y},L_{z}\leq 64$. We calculate the resistivity
in the three
directions by applying a small probe current and evaluating the average voltage. For example, for
the resistivity $\rho_{\mu}$ in the $\mu-$direction we drive the system with a small current
$I_{\mu}$ ($I_{\mu}=0.01I_{c}^{\mu}$) and measure the voltage 

\begin{equation}
V_{\mu}=\frac{\hbar}{2e} \langle \frac{d}{dt}(\varphi ^{i+\mu}-\varphi ^{i})\rangle
\end{equation}
Then $\rho_{\mu}=V_{\mu}/I_{\mu}$.

\subsection{Transport properties}

The results for the resistivity calculated with periodic boundary conditions
(PBC) along the field direction and free boundary conditions (FBC)\ in the
other directions are shown in Fig.(3). These results correspond to a sample with $f=1/12$, $\gamma
^{2}=20$, $L_{x}=L_{y}=L_{z}=30$, $\Delta=.05$ and $N=50,000$. We found similar results for systems
with $f=1/48,1/24,1/6$, $\gamma ^{2}=5,10,25$ and the same sizes and disorder as the ones indicated above. The
general behavior is qualitatively similar to the experimental data obtained by Grigera \textit{et al.}
shown in the inset for comparison.

\begin{figure}[tbp]
\centerline{\epsfxsize=8.5cm \epsfbox{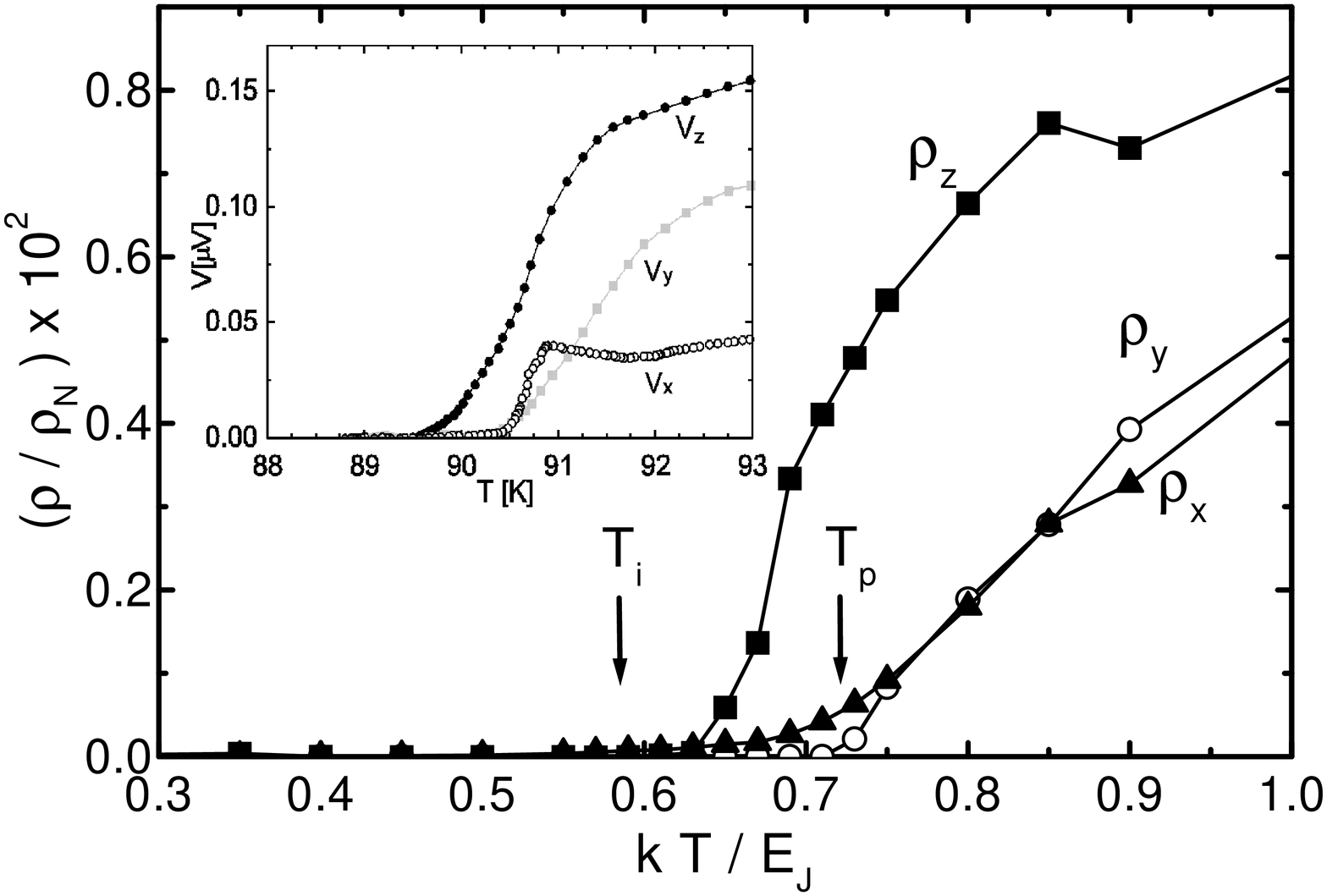}}
\caption{Resistivities in the three directions for a system with FBC in $x$ and $z$
directions and PBC in the field direction. The parameters are $f=1/12$, $\gamma ^{2}=20$, $\Delta=.05$,
$N=50,000$ and $L_{x}=L_{y}=L_{z}=30$. $\rho_{y}$ is the resistivity parallel to
the field, $\rho_{x}$ is the perpendicular to the field but parallel to the ab-planes
 and $\rho_{z}$ is perpendicular both to the planes and the field. In
the inset experimental results of Ref.[11] are shown.}
\label{fig3}
\end{figure}

The response of the system when the external
current is perpendicular to the planes (and the Lorenz force is parallel to
them) is given by $\rho _{z}$. For this geometry, the pinning due to the planes is unimportant and
for temperatures higher than a characteristic temperature $T_{i}$ the
resistivity $\rho _{z}$ increases rapidly. The stable phase for $T>T_{i}$ corresponds to
a liquid phase and the rapid increase of $\rho _{z}$ is an indication of the
high mobility of vortices parallel to the planes. The
transition from the low temperature phase to the high temperature liquid
phase is continuous. These results, obtained in the finite system, are not
conclusive on whether the change of behavior observed at $T_{i}$ is a second
order phase transition or a crossover between two different regimes. In finite systems,
thermal activation is observed down to low temperatures and
finite size scaling is necessary in order to characterize the transition.
Based on previous results and the discussion of next section, we will refer
to this behavior as continuous transition and to zero resistivity when the
noise in the voltage is larger than its mean value, corresponding typically
to $\rho _{\mu}<R_{\mu}\times 10^{-5}$, where $\mu=x,y,z$.

When the external current is applied along the $x-$direction, the Lorenz
force is perpendicular to the planes and for temperatures $T\gtrsim T_{i}$
the in-plane pinning dominates the vortex dynamics resulting in a small $%
\rho _{x}$. The resistivity $\rho _{x}$ decreases strongly near a characteristic
temperature $T^{*}>T_{i}$ and has a tail that extends down to $T_{i}$. This
behavior is obtained for different values of the anisotropy and magnetic
field, it qualitatively reproduces the experimental observation and
resembles the prediction of the smectic phase theory. In this theory, as the
temperature is lowered, the system undergoes a transition to a smectic phase
at a temperature $T_{s}$. In the temperature interval $T_{i}<T<T_{s}$ the
system develops long ( or quasi-long) range order in the direction
perpendicular to the planes and the resistivity shows, in the vicinity of
the smectic temperature $T_{s}$, a critical behavior of the form $\rho
_{x}(T)-$ $\rho _{x}(T_{s})\propto |T-T_{s}|^{1-\alpha}$ where $\alpha$ is
the specific heat critical exponent. At the temperature $T_{i}$, long range
order in the direction parallel to the planes appears giving rise to a
crystalline structure. Since in finite size systems the resistivity is not
necessarily a good quantity to characterize a phase transition (in particular
it will not show an infinite slope) we search for some signature of a
smectic phase by analyzing the structure of the liquid for temperatures $%
T\gtrsim T_{i}$ corresponding to the regime where $\rho _{x}$ shows a tail.
The details of the structure factors calculated with PBC at
different temperatures are presented in next section.

Finally, we have also calculated the resistivity $\rho _{y}$ corresponding
to a transport current along the field direction for which there is zero
Lorenz force. In this geometry, the resistivity goes to zero at a
temperature $T_{p}>T_{i}$. This behavior is due to finite size effects;
dissipation occurs when vortices entangle to form a structure that
percolates in the directions perpendicular to the external field. This
happens at a temperature $T_{p}$ that is sensitive to the thickness ($L_{y}$) of the
sample, the thicker the sample, the lower is $T_{p}$ . Previous studies\cite{jagla} of
the resistivity in the direction of the field show that in the thermodynamic
limit $T_{p}$ coincides with $T_{i}$ as experimentally observed and we
expect the resistivity along the three directions to vanish at the same
temperature.

\begin{figure}[tbp]
\centerline{\epsfxsize=8.5cm \epsfbox{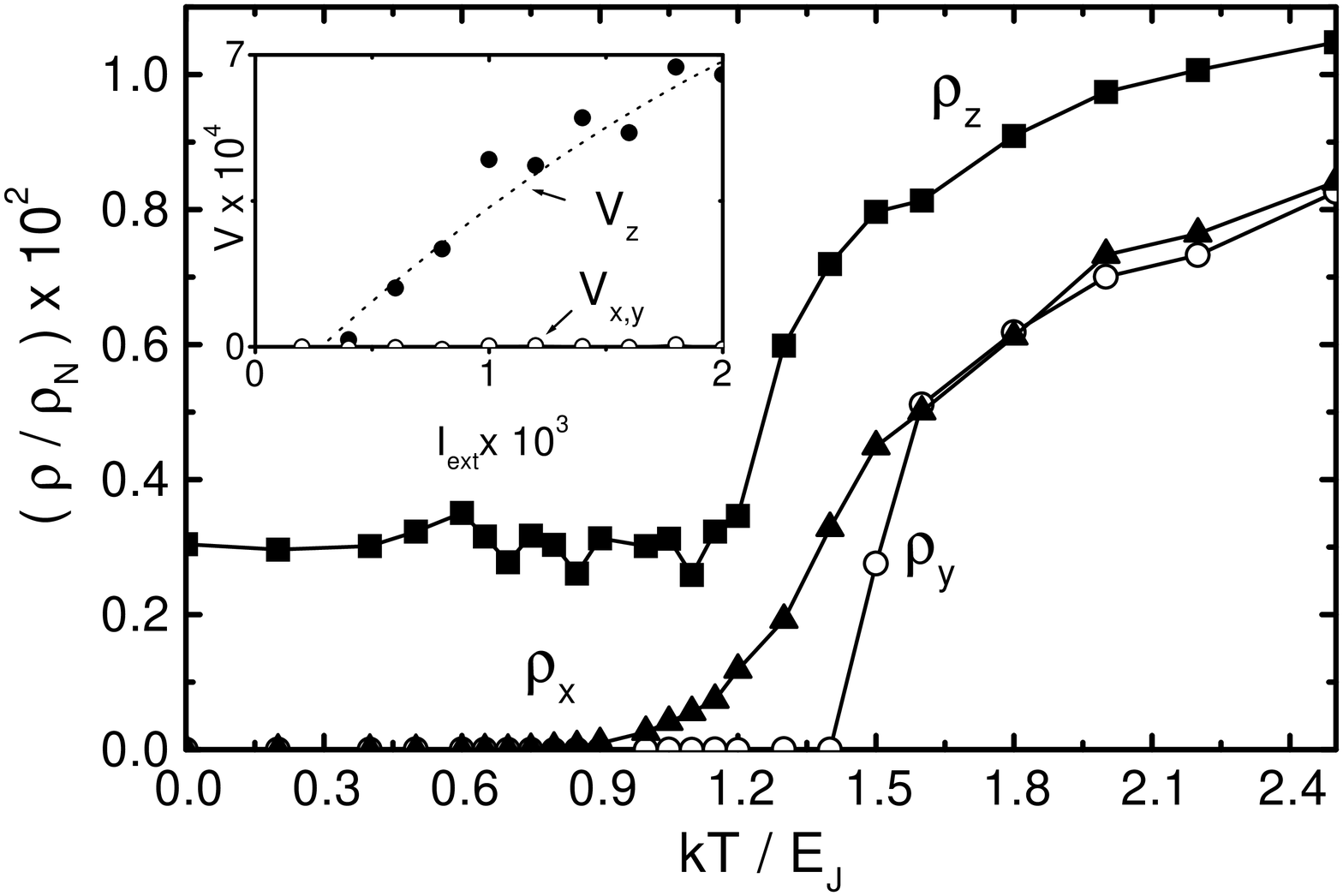}}
\caption{Resistivities in the three directions with PBC for a
system with $f=1/24$, $\gamma ^{2}=9$, $L_{x}=L_{z}=48$ and $L_{y}=8$. $\rho_{x}$,
$\rho_{y}$ and $\rho_{z}$ are defined in Fig.(3). In the inset the I-V
characteristics for the current along the three directions are shown.}
\label{fig4}
\end{figure}

In order to discard boundary effects due to surface barriers, we have also
calculated the resistivities and structure factors using PBC in the three directions.
Results for the resistivity in a highly anisotropic sample with weak disorder are shown
in Fig.(4). For this case, the parameters are $f=1/24$, $\gamma^{2}=9$, $L_{x}=L_{z}=48$, $L_{y}=8$ $\Delta=0$ and
$N=30,000$. We have also done some simulations with PBC in the three directions for $f=1/72,1/48,1/8$,
$\gamma^{2}=1 - 40$ and $N=10^{4} - 10^{5}$, and we found similar results.
While $\rho _{x}$ and $\rho _{y}$ behave essentially as in the previous
case, as the temperature decreases $\rho _{z}$ saturates at a value
different from zero, which increases with anisotropy. The I-V characteristics for the current along the three
directions at low temperatures are shown in the inset and indicates a very
small value of the critical current in the $z-$direction. The saturation of
the resistivity at low temperatures is due to the fact that the run was done
with a value of the transport current $j_{ext}$ larger than the critical
current $j_{c}$ and the results correspond to a flux flow regime. A
systematic study of the resistivities with $j_{ext}<$ $j_{c}$ requires much
more statistics and generates larger numerical errors. In any case, these
results show two important points: first, for the systems under
consideration, $\rho _{x}$ and $\rho _{y}$ are not very sensitive to the
boundary conditions, and second, when vortices flow parallel to the planes,
pinning is very weak. In the case of FBC the surface barriers generate some
extra pinning that increases the critical current. In the thermodynamic
limit, any rigid structure is pinned by impurities and surface effects and
in this sense, it may be more appropriate to compare the case of FBC with experiments if
results with $j_{ext}<$ $j_{c}$ are not available for the case of PBC. As a side comment,
notice that the noise in $%
\rho _{z}$ disappears at a temperature of the order of $T_{i}$ at which $%
\rho _{y}$ vanishes. A simple interpretation of this effect is that in the
flux flow regime an ordered structure generates less noise than a liquid.
However a systematic study of the noise, that could give information on the
nature of the different phases, is needed to reach definitive conclusions.

\subsection{Structure factor}

In this section we present results for the structure factor calculated for
different temperatures. At high temperatures, in the liquid phase, the
system is highly disordered and presents vortex loop excitations. An instantaneous picture
of the vortices crossing a plane perpendicular to the external field ($%
y=L_{y}/2$) is shown in Fig.(5a) for the same system as Fig.(4).

\begin{figure}[tbp]
\centerline{\epsfxsize=8.5cm \epsfbox{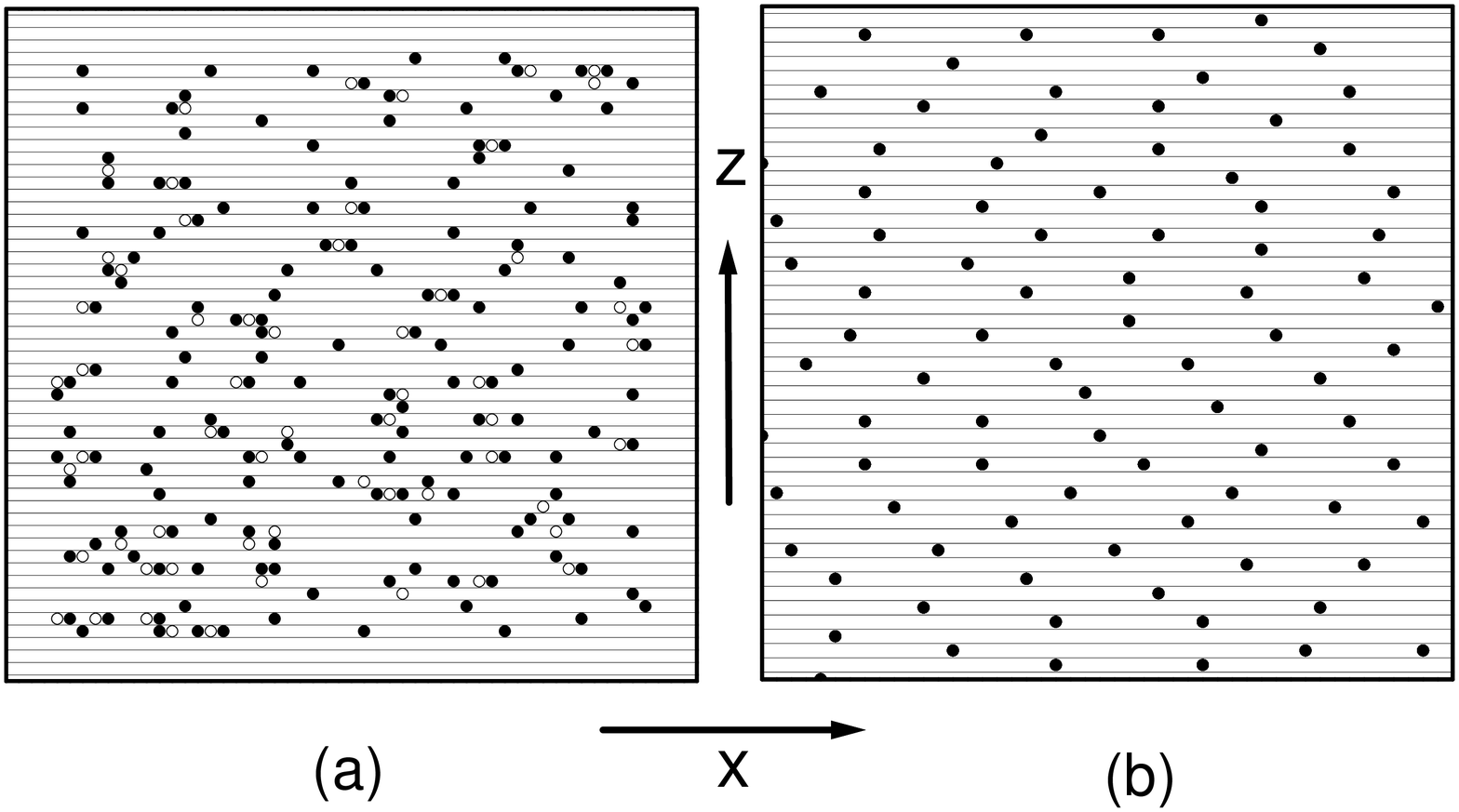}}
\caption{Instantaneous configuration of the vortices in a plane perpendicular to the field for the system of
Fig.(4). Black dots are vortices and white ones are antivortices.
In (a) the temperature is $\frac{kT}{E_{j}}=1.1$ and in (b) $\frac{kT}{E_{j}}=0$.}
\label{fig5}
\end{figure}

The vortex-antivortex pairs are small
loops confined between two $ab-$planes and cut by the $y=L_{y}/2$ plane.
As the temperature decreases the system evolves towards an ordered solid
structure. It is known that, due to numerical limitations, in three
dimensions it is very difficult to cool the system into an ordered lattice.
The type of structures obtained by slowly cooling the system is shown in Fig
(5b). We obtain this kind of structures for systems with $f \leq 1/12$. For higher magnetic fields, a triangular
vortex lattice was found like the one predicted by Ref.[3].

Although the obtained structure is very disordered, the tendency to form
vortex chains (at approximately $45^{0}$ in the figure) can be observed. This tendency is reflected by
the structure factor $S({\mathbf q}_{\perp},y)$ defined as\cite{Teitel}:

\begin{equation}
S({\mathbf q}_{\perp },y)=\frac{1}{N^{2}}\left| \sum_{j}\eta _{j}e^{i{\mathbf q}_{\perp}.{\mathbf
r}_{j}} \right| ^{2}  \label{sq}
\end{equation}
where $N$ is the total number of vortices, $\eta _{j}$ is the\textit{\
vortex charge}: 1 for vortices and -1 for antivortices, and $\mathbf{r}_{j}$
are the vortex coordinates in the direction perpendicular to the external
field at a plane with coordinate $y$. We define $S(\mathbf{q}_{\perp })$ as
the average over $y$ of $S({\mathbf q}_{\perp},y).$

In Fig.(6) the high and low temperature structure factor for the system of Fig.(5) are shown.

\begin{figure}[tbp]
\centerline{\epsfxsize=8.5cm \epsfbox{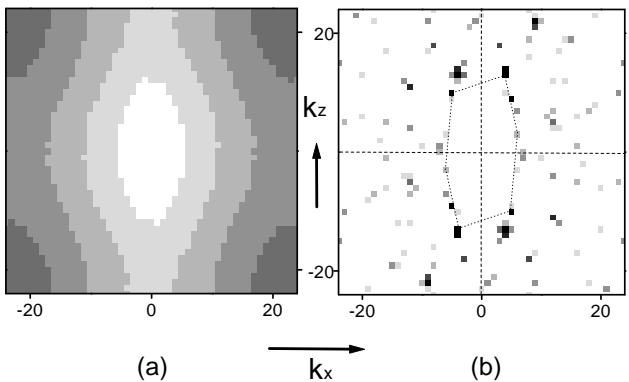}}
\caption{Structure factor of the configurations of Fig.(5). (a) and (b) correspond to the same
temperatures showed in that figure.}
\label{fig6}
\end{figure}

At high temperatures only a background of the form $\cos (q_{x}a)$, where $a$
is the lattice parameter of the junction network, is obtained. This
background is due to the presence of small loops which in general are
confined between to consecutive $ab-$planes, \textit{i.e. }vortex-antivortex
pairs which are oriented in the $x$-direction and bound at a distance $a$.
At low temperatures, the background disappears and well defined peaks are
observed. The peaks in the structure factor correspond to a rotated vortex
lattice with a small angle $\theta $. In Fig.(6b) we draw one of the corresponding rotated hexagonal
reciprocal space unit cell. We also see that there are peaks corresponding to the same lattice
structure reflected in $180 ^{0}$ with respect to the $z-$axis. The coexistence of these two
reflected structures could explain the chain-like ordering observed in real space configurations
(Fig.(5b)).
For this particular value of the
external field, we follow the evolution of the structure factor as the
system is cooled and observe some indications of a smectic phase at
intermediate temperatures. A sequence of the structure factor at
intermediate temperatures is shown in Fig.(7).

At temperatures $T\gtrsim T_{i}$, where the resistivity $\rho _{x}$ presents
a small tail, a peak at ${\mathbf q}=(0,q_{y}^{0})$ is clearly observed. This
peak indicates the presence of a smectic phase. As the temperature is
lowered from high temperatures, first the intensity of the smectic peak
increases, goes through a maximum and then decreases as the crystalline
peaks corresponding to a rotated structure increase. The wave vector $%
q_{y}^{0}$ may also be weakly temperature dependent: The first smectic peak
observed has $q_{y}^{0}\simeq  Q/3$ ($Q=2\pi /s$) and as the temperature is
lowered it shift towards $ Q/2$. In Fig.(8) the temperature
dependence of the smectic peak and one of the crystalline peaks are
shown. The oscillations in the amplitude of the smectic peak are probably due
to finite size effects, since only some discrete values of  $q_{y}^{0}$ are
consistent with the PBC. At temperatures $T\sim T_{i}$ both the smectic and
the crystalline peaks coexist, indicating the coexistence of the two
phases.

\begin{figure}[tbp]
\centerline{\epsfxsize=8.5cm \epsfbox{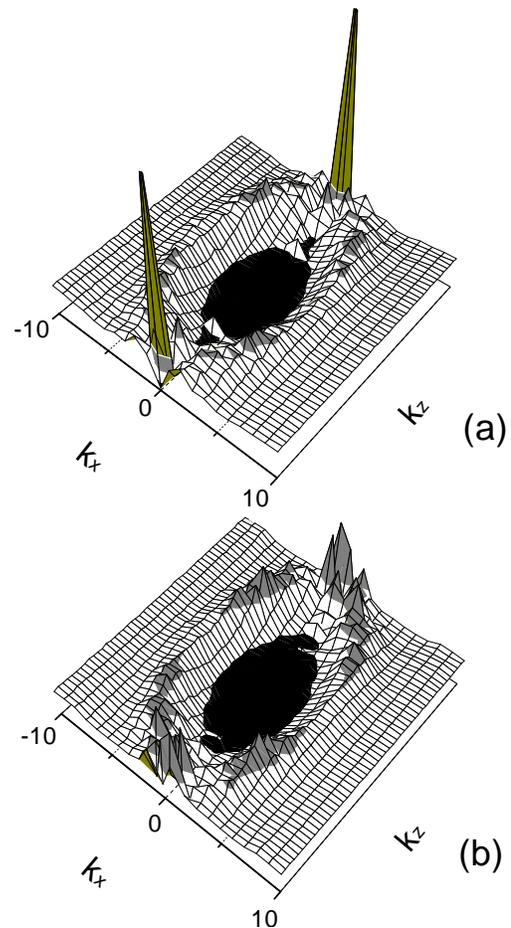}}
\caption{Structure factor for the system of Fig.(5) at intermediate temperatures. A smectic peak can be
observed at $q_{y}^{0}\simeq Q/2$ at $T=0.53$ (a), which decreases at a lower temperature
T=0.30 (b).}
\label{fig7}
\end{figure}

Since the symmetry group of the smectic phase is not a subgroup of the one
corresponding to the symmetry of the rotated lattice, we expect a first order
transition between these two phases. This is consistent with the observation
of a coexistence of the two phases in the numerical simulations.

For other values of the parameters studied, ($f=1/24$ and $\gamma ^{2}=2.56, 25$), for which the low
temperature phase corresponds to a rotated crystal with a larger rotation
angle, we do not observe indications of a smectic order in the liquid
phase. In these cases, our preliminary results indicate a continuous
transition from the liquid to the frozen state without any precursor of long
or quasi long range order in the direction perpendicular to the planes.

\begin{figure}[tbp]
\centerline{\epsfxsize=8.5cm \epsfbox{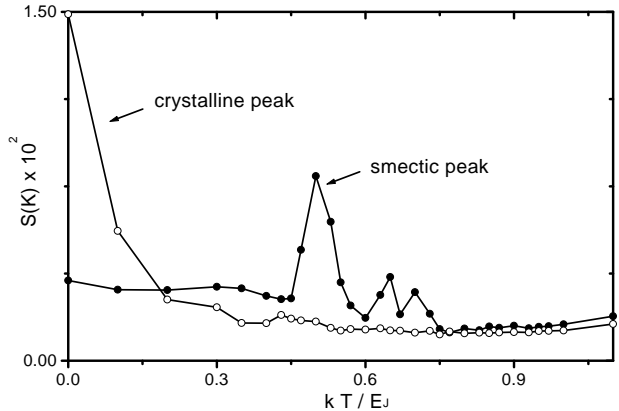}}
\caption{Temperature dependence of the intensity of the smectic peak described in the text (open symbols) and a
crystalline peak, $(q_{x}^{0},q_{y}^{0})\simeq (Q/12,Q/4$) (filled
symbols).}
\label{fig8}
\end{figure}

\section{SUMMARY AND DISCUSSION}

We have studied the problem of the low temperature structure, the
thermodynamic and transport properties of vortices when the external field
is applied parallel to the CuO planes in high $T_{c}$ superconductors. A
simple analysis based on the London theory in the presence of a periodic
potential indicates the possibility of a variety of commensurate structures
that essentially correspond to anisotropy-distorted Abrikosov lattices
rotated to match the periodic potential. For low fields a large number of
degenerate structures corresponding to different rotation angles are
obtained. Using parameters like those typically used to describe YBaCuO, $s = 10\AA $ and a mass anisotropy
$m_{1}/m_{3}=50$, for magnetic induction between $5$ and $6$ Tesla, we obtain about seven rotation angles
which generate commensurate structures.

The numerical simulations in a XY-model clearly show the tendency to form
rotated structures. Our results shows structures with grains of twin phases.

The transport properties qualitatively reproduce the experimental results.
In particular,  $\rho _{x}$ shows a rapid decrease and a tail that extends
down to $T_{i}$. This behavior is similar to the predictions of the smectic
phase theory although transport properties seem not to be enough to prove
the existence of this phase at an intermediate temperature range.

When the rotated angle of the low temperature structure, which depends both
on the value of the external field and the anisotropy, is small, we observe
well defined peaks corresponding to a smectic phase. The vector \textbf{Q }%
and\textbf{\ }the amplitude of the smectic peak are temperature dependent and
a first order transition is expected from the smectic to the crystalline
phases. For the parameters that give a larger rotation angle of the
crystalline phase, we observe a direct evolution of a system from the
anisotropic liquid to the vortex lattice without any indication of an
intermediate smectic phase.

\acknowledgements
We acknowledge stimulating discussions with E.A. Jagla and E. Osquiguil, and thank S. Grigera for the data of
Fig.(3). One of us (M.F.L.) acknowledge support from program FOMEC. We also acknowledge financial support from
CONICET, CNEA, ANPCyT and Fundaci\'{o}n Antorchas.


\begin{references}
\bibitem{crabtree}  G. Crabtree and D. Nelson, Physics Today \textbf{50},
38, April 1997.

\bibitem{blatter}  G. Blatter \textit{et al.}, Rev. Mod. Phys. \textbf{66}
1125 (1994).

\bibitem{bula} L.N Bulaevskii and J.R. Clem, Phys. Rev. B \textbf{44} 10234 (1991).

\bibitem{ivlev}  B. I. Ivlev and L. J. Campbell, Phys. Rev. B \textbf{47}
14514 (1993).

\bibitem{Kwok}  W.K. Kwok \textit{et al., }Phys Rev. Lett. \textbf{72}, 1088
(1994).

\bibitem{bal1}  L. Balents and D.R. Nelson, Phys. Rev Lett. \textbf{73},
2618 (1994).

\bibitem{bal2}  L. Balents and D.R. Nelson, Phys. Rev. B \textbf{52}, 12951
(1995).

\bibitem{Hu}  X. Hu and M. Tachiki, Phys. Rev. Lett. \textbf{80} 4044 (1998).

\bibitem{Olson}  P. Olsson and P. Holme, cond-mat 9907118.

\bibitem{CDK}  L.J. Campbell \textit{et al.}, Phys. Rev B \textbf{38}, 2439
(1988).

\bibitem{Grig}  S.A. Grigera \textit{et al.}, Phys. Rev. B \textbf{59},
11201 (1999).

\bibitem{hu2}  Hu and Tachiki (unpublished).

\bibitem{ivlev2}  B.I. Ivlev, N.B. Kopnin and V.L. Pokrovsky, J. Low Temp.
Phys. \textbf{80}, 187 (1990).

\bibitem{prokov}  V.L Pokrovsky and A.L Talapov, Phys Rev. Lett \textbf{42},
65 (1979); Sov. Phys. JETP \textbf{51}, 134 (1980).

\bibitem{martin}  P. Martinolli \textit{et al.,} Helvetica Physica Acta,
Vol. \textbf{55}, 655 (1982).

\bibitem{dolan}  G. J. Dolan F. Holtzberg, C. Field and T. R. Dinger, Phys.
Rev. Lett. \textbf{62} 2184 (1989).

\bibitem{Mingo}  D. Dom\'{\i}nguez, N. Gr{\o}nbech-Jensen and A.R. Bishop, Phys. Rev. Lett. \textbf{75},
717 (1995); Phys. Rev. Lett. \textbf{75}, 4670 (1995); Phys. Rev. Lett. \textbf{78}, 2644 (1997). 

\bibitem{jagla}  E.A. Jagla and C.A. Balseiro, Phys. Rev. B \textbf{53},
R538 (1996); Phys. Rev. B \textbf{53}, 15305 (1996).

\bibitem{hetzel} R.E. Hetzel \textit{et al.}, Phys. Rev. Lett. \textbf{69} 518 (1992).

\bibitem{li}  Y.-H. Li and S. Teitel, Phys. Rev. Lett. \textbf{66}, 3301 (1991).

\bibitem{hu3}  X. Hu, S. Miyashita and M. Tachiki, Phys. Rev. Lett. \textbf{79} 4044 (1997).

\bibitem{Teitel}  Y.-H. Li and S. Teitel, Phys. Rev. B \textbf{47}, 359
(1993).

\bibitem{tachiki}  M. Tachiki and S. Takahashi, Solid State Commun. \textbf{%
70}, 291 (1989).

\end{references}
\end{document}